**Identifying the most predictive risk factors for future cognitive impairment among elderly Chinese**

Collin Sakal, MSc[1]; Tingyou Li[1]; Juan Li, PhD[2]; Xinyue Li, PhD[1*]

**Affiliations**

1. School of Data Science, City University of Hong Kong, Hong Kong SAR, China

2. Center on Aging Psychology, Key Laboratory of Mental Health, Institute of Psychology, Chinese Academy of Sciences, Beijing, China

***Address correspondence to:** Xinyue Li, PhD

Xinyue Li, PhD

School of Data Science, City University of Hong Kong, Hong Kong SAR, China

Postal address: 83 Tat Chee Avenue, Lau-16-224, Kowloon Tong, Hong Kong SAR, China

Telephone number: (+852)34422180

Email: xinyueli@cityu.edu.hk

**Word Count:** 2999

## Key Points

**Question:** Which risk factors most accurately predict future cognitive impairment among elderly Chinese and are there any population subgroups where predictions tend to be less accurate?

**Findings:** Demographics, cognitive tests, and instrumental activities of daily living had the most predictive power out of nine risk factor groups. The most predictive risk factors and three existing prediction models had significantly higher AUCs among female and uneducated elderly compared to male and educated elderly.

**Meaning:** Greater efforts are needed to make equally effective risk predictions for future cognitive impairment across different subpopulations in China.

**Abstract**

**Importance**: In China, the societal burden of cognitive impairments continues to increase as the country ages, but our knowledge remains limited regarding how to accurately predict future cognitive impairment at the individual level for preventative interventions. Identifying the most predictive risk factors and socioeconomic groups where predictions are less accurate would provide a foundation for developing targeted prediction models that can identify elderly at high risks of future cognitive impairments.

**Objectives**: To quantify how well demographics, instrumental activities of daily living, activities of daily living, cognitive tests, social factors, psychological factors, diet, exercise and sleep, chronic diseases, and three recently published prediction models predict future cognitive impairments in the general Chinese population and among male, female, rural, urban, educated, and uneducated elderly.

**Design**: The Chinese Longitudinal Healthy Longevity Survey (CLHLS) is a prospective cohort study of elderly Chinese from 23 provinces. Individual information from the 2011 CLHLS survey was used to predict if participants would become cognitively impaired by follow-up in 2014.

**Setting**: Population-based.

**Participants**: 4047 CLHLS participants 60 years of age or older without cognitive impairments at baseline were included.

**Main Outcome:** Cognitive impairment was identified through the Chinese language version of the Mini Mental State Examination (MMSE). Predictive ability was quantified using the AUC, sensitivity, and specificity across 20 repeats of 10-fold cross validation where the target variable was an indicator of cognitive impairment 3 years from the baseline survey.

**Results**: A total of 337 (8.3%) elderly Chinese became cognitively impaired by the follow up survey. The risk factor groups with the most predictive ability in the general population were demographics (AUC, 0.78, 95% CI, 0.77-0.78), cognitive tests (AUC, 0.72, 95% CI, 0.72-0.73), and instrumental activities of daily living (AUC, 0.71, 95% CI, 0.70-0.71). Demographics, cognitive tests, instrumental activities of daily living, and all three re-created prediction models had significantly higher AUCs when making predictions among women compared to men and among the uneducated compared to the educated. Dietary factors, which have yet to be included in prediction models in China, had more predictive power (AUC, 0.59, 95% CI, 0.58-0.60) than activities of daily living (AUC, 0.57, 95% CI, 0.56-0.57), psychological factors (AUC, 0.58, 95% CI, 0.57-0.59), and chronic diseases (AUC, 0.53, 95% CI, 0.52-0.53).

**Conclusion and relevance**: This study suggests that demographics, cognitive tests, and instrumental activities of daily living are the most useful risk factors for predicting future cognitive impairment among elderly Chinese. However, the most useful risk factors and existing models have lower predictive power among male, urban, and educated elderly Chinese. More efforts are needed to ensure that equally accurate risk assessments can be conducted across different socioeconomic groups in China.

# Introduction

China's aging population has led cognitive impairments to become increasingly burdensome to society[1,2]. In 2020 more than 68 million elderly Chinese had mild cognitive impairments, dementia, or Alzheimer's[3]. The economic and social burden of cognitive impairments has led to calls for improving risk assessments and prioritizing early diagnoses[1,3]. Given China's limited number of geriatric psychiatrists, researchers have turned to developing prediction models to identify elderly at risk of cognitive impairments for preventative interventions[4-8]. However, no study has compared the predictive ability of known risk factors side by side, and our understanding of which factors are most useful for developing prediction models is limited. Furthermore, population characteristics vary widely across China, but it is unknown which risk factors are most predictive in different socioeconomic groups, and existing prediction models have primarily been tested in the general population alone. To understand how to best predict future cognitive impairments, and to develop more targeted prediction models for population subgroups, the predictive ability of known risk factors and existing prediction models must be quantified and compared across different subsets of the Chinese population.

Demographics, instrumental activities of daily living, chronic diseases, and certain cognitive tests are all associated with future cognitive impairment[3,9], but which are the most useful in a predictive modeling context remains uncertain. As previous work has shown, significant odds ratios reported in association studies do not guarantee that such risk factors will result in clinically viable prediction models[10]. Furthermore, the prevalence of cognitive impairments in China varies between male, female, rural, urban, educated, and uneducated elderly[3,11-13], but no foundational work has been published detailing which risk factors are most useful for predicting future cognitive impairments in each of these subpopulations. Hence, while we have a reasonable understanding of which characteristics make a person more likely to develop cognitive impairments, we have a poor understanding of which parts of a person's health profile most accurately predict their likelihood of developing cognitive impairments.

In addition to known risk factors, it is unclear if existing prediction models for future cognitive impairment are equally accurate across different socioeconomic groups in China. Several published models have reported AUCs greater than 0.80 in their development cohorts[5,7,8], but each has only been tested on the general population. Examining the predictive ability of existing models across population subsets would allow us to identify where more efforts are needed to improve risk assessments for cognitive impairment, further our understanding of which subpopulations are more difficult to conduct risk assessments within, and provide a more thorough evaluation of existing prediction models than has been reported previously.

In this study we quantified the ability of nine risk factor groups and three existing models to predict future cognitive impairment among elderly Chinese. We examined how well demographics, instrumental activities of daily living (IADLs), activities of daily living (ADLs), cognitive tests, social factors, psychological factors, diet, exercise and sleep, chronic diseases, and three recently published models predict future cognitive impairments in the general population and among male, female, rural, urban, educated, and uneducated elderly. To our knowledge, this study is the first to comprehensively compare the ability of known risk factors to predict future cognitive impairment, and the first seeking to identify which subsets of the Chinese population need greater attention to improve the accuracy of risk assessments.

**Methods**

**Data Source and Study Design**

The Chinese Longitudinal Healthy Longevity Survey (CLHLS) is a prospective cohort study of elderly Chinese that contains information on demographics, cognitive function, lifestyle factors, chronic diseases, and more[14,15]. The CLHLS began in 1998 and follow-up surveys have been conducted every 2-3 years since. Data include elderly from 23 of China's provinces that together make up 85% of the country's total population. The CLHLS received ethics approval from the Duke University Institutional Review Board (Pro00062871) and Peking University's Biomedical Ethics Committee (IRB00001052–13,074). Written informed consent was given by all participants prior to the survey interviews.

We used the 2011 and 2014 CLHLS surveys in our study. Baseline characteristics were gathered from the 2011 survey and used to predict if an individual became cognitively impaired by 2014. CLHLS participants under the age of 60 or with cognitive impairments at baseline were excluded. Sample size calculations were conducted following Riley's methodology for multivariable prediction models[16]. This study is presented following the TRIPOD guidelines (eTable 12 in the Supplement) where appropriate[17,18].

**Measuring Cognitive Function**

Cognition was assessed through the Chinese language version of the Mini Mental State Examination (MMSE)[19]. MMSE scores range from 0 to 30 with lower scores indicating worse cognitive function. We adopted education specific cutoffs that have been previously validated in the elderly Chinese population to indicate cognitive impairment[20]. Those with no formal education and MMSE scores less than 18 were labeled as cognitively impaired, as were those with 1-6 years of education with scores less than 21, as were those with more than 6 years of education with scores less than 25[20].

**Risk Factor Groups**

Nine groups containing known risk factors for cognitive impairment were considered in this study: demographics, activities of daily living (ADLs), instrumental activities of daily living (IADLs), cognitive tests, social factors, psychological factors, exercise and sleep, diet, and chronic diseases. The risk factor groups were chosen by selecting parts of a person's health profile previously found to be associated with developing cognitive impairments[9]. Each group is briefly described below, and a complete list of the variables in each group can be found in eTable 1.

**Demographics**

The demographics group contained each individual's age, sex, years of education, household income, marital status, and residence location (city, town, rural).

**Activities of Daily Living**

The activities of daily living (ADL) group included each person's ability to bathe, get dressed, use the toilet, get in and out of bed, control urination and bowel movements, and eat food.

**Instrumental Activities of Daily Living**

Instrumental activities of daily living (IADLs) cover tasks that require thinking, organizational, and physical independence. The IADL group included an elderly person's ability to visit neighbors, go shopping, cook, wash clothes, walk continuously for 1km, lift a bag of groceries, crouch and stand up, and take public transportation.

**Cognition**

The cognition group included scores from each sub-section of the MMSE: orientation, naming, registration, calculation, attention, recall, and language. Scores from each section were included as separate variables.

**Social Factors and Hobbies**

The social factors group included whether a person grows vegetables, gardens, reads newspapers and books, looks after pets or animals, plays cards or mahjong, and whether they participate in social activities.

**Psychological Factors**

The following factors, which primarily relate to depression and anxiety, were included in the psychological factors group: whether a person is generally optimistic, keeps their belongings organized, is generally anxious, is often lonely, makes decisions independently, feels useless with age, was happier when they were younger, and felt sad for more than two consecutive weeks over the past year.

**Exercise and sleep**

The exercise and sleep group included whether someone currently exercises, whether they used to exercise, as well as the self-reported duration and quality of sleep.

**Diet**

The dietary group contained information on each person's staple food, if they eat fresh fruits and vegetables, the main flavor of the dishes they cook, how frequently they consume meat, fish, eggs, sugar, and tea, if they consume alcohol, the type of alcohol they consume, and the frequency of alcohol consumption.

**Chronic Diseases**

In the chronic diseases group we included the presence or absence of hypertension, diabetes, heart disease, blood disease, and cardiovascular disease.

**Re-creating Existing Prediction Models**

Prediction models were selected based on the following criteria: the model was developed for use in China, was reproducible using the CLHLS, and had an AUC of >0.75 during development. We selected three models published in Zhou (2020)[8], Hu (2021)[5], and Wang (2022)[7]. Each model was developed for use in the general Chinese population and showed excellent predictive performance (AUC > 0.80) during development. A complete description of the predictors in each model can be found in eTable 2.

**Statistical Analysis**

All analyses were performed using the R Statistical Software (version 4.0.5, R Foundation for Statistical Computing, Vienna, Austria). Predictive ability was quantified using the area under the receiver operating characteristic curve (AUC), sensitivity, and specificity. We assessed the predictive ability of each risk factor group through twenty repeats of 10-fold cross validation, which has been recommended to obtain optimism corrected performance metrics for prediction models[21]. This resulted in 200 training and 200 validation sets respectively. Ordinal variables were integer encoded, and non-ordinal categorical variables were dummy encoded. Missing values were imputed on each training and validation set separately using K-Nearest Neighbors (KNN) imputation[22]. During each iteration of cross-validation the data were split into training and validation sets before nine logistic regression models, one for each risk factor group, were fit to the training data. Thereafter, each model was used to make predictions on the validation set for the general population and six sub-populations: male, female, rural, urban, uneducated, and educated elderly. The same procedure was also followed for evaluating Hu, Zhou, and Wang's prediction models. Average AUCs and accompanying 95% confidence intervals were calculated from the validation set observations. Sensitivity and specificity curves, one from each validation set, were also plotted for the risk factor group models.

## Results

Given a binary outcome, a population-level prevalence of 0.20, a conservatively estimated Cox-Snell $R^2$ of 0.09, and 24 predictors in the largest risk factor group, the sample size required for this study was determined to be 1065 with 213 events. After excluding CLHLS participants with cognitive impairments at baseline and those under the age of 60, a cohort of 4047 elderly Chinese, of which 337 developed cognitive impairment, were included. The average age of the cohort was 79.8 and 2037 (50%) were men. The group that developed cognitive impairments was older at baseline (89.1 versus 79.0 years) with a lower average baseline MMSE score (25.1 versus 27.7). A full description of the cohort's characteristics can be found in Table 1 and the distribution of variables in each risk factor group can be found in eTables 3-11.

As shown in Figure 1A and Table 2, the demographics group had the best predictive ability in the general population (AUC, 0.78, 95% CI, 0.77-0.78) followed by cognitive tests (AUC, 0.72, 95% CI, 0.72-0.73) and instrumental activities of daily living (AUC, 0.71, 95% CI, 0.70-0.71). Social factors had moderate predictive ability (AUC, 0.67, 95% CI, 0.66-0.68) while diet, psychological factors, exercise and sleep, ADL, and chronic diseases all had average AUCs less than 0.60. Demographics, cognitive tests, and IADLs also had the best sensitivity/specificity tradeoffs as shown in Figure 2. By contrast, the sensitivity and specificity curves for the chronic diseases group showed that such risk factors only sometimes resulted in better-than-random predictions.

Demographics, cognitive tests, and IADLs also had the highest average AUCs when making predictions for men and women, though predictive ability varied between the two sexes. The demographics group had a higher average AUC when making predictions in women compared to men (0.81, 95% CI, 0.80-0.82 versus 0.72, 95% CI, 0.71-0.73), as did the IADLs group (0.72, 95% CI, 0.71-0.73 versus 0.66, 95% CI, 0.65-0.67) and the cognition group (0.72, 95% CI, 0.71-0.73 versus 0.70, 95% CI, 0.69-0.71). The dietary group had a significantly higher AUC when making predictions among men (0.61, 95% CI, 0.60-0.62) compared to women (0.57, 95% CI, 0.56-0.58). No significant differences were observed for the social factors and hobbies group, and all other remaining groups has AUCs less than 0.60 for both men and women. Full results can be found in Figure 2B and Table 2.

Demographics had a significantly better predictive ability when making predictions among rural elderly (AUC, 0.80, 95% CI, 0.79-0.81) compared to urban dwelling elderly (AUC, 0.74, 95% CI, 0.73-0.75). Similarly, IADLs showed a higher average AUC among rural dwellers (AUC, 0.73, 95% CI, 0.72-0.73) compared to urbanites (AUC, 0.68, 95% CI, 0.67-0.69). As shown in Figure 1C and Table 2, significantly higher AUCs among rural elderly were also observed for the diet, psychological factors, and chronic diseases groups.

Figure 2D shows that most risk factor groups had significantly higher AUCs when making predictions among the uneducated compared to the educated. The only exceptions were the ADL and exercise and sleep groups. Among the uneducated, demographics, cognitive tests, and IADLs had AUCs of 0.79 (95% CI, 0.78-0.79), 0.72 (95% CI, 0.72-0.73), and 0.71 (95% CI, 0.70-0.71) respectively. When making predictions among the educated, demographics, cognitive tests, and IADLs had average AUCs of 0.73 (95% CI, 0.72-0.74), 0.64 (95% CI, 0.63-0.66), and 0.64 (95% CI, 0.62-0.65).

The existing prediction models re-created in this study all had good predictive ability in the general population. Each model had an average AUC of 0.80 (95% CI, 0.80-80.1). However, every model had significantly higher AUCs when making predictions in women compared to men, in rural elderly compared to urban elderly, and among the uneducated compared to the educated. Complete results can be found in Table 2 and Figure 3.

## Discussion

In this study we quantified the ability of nine risk factor groups and three prediction models to predict future cognitive impairment in the Chinese population and six population subsets: male, female, rural, urban, uneducated, and uneducated elderly. In the general population, the risk factor groups with the best predictive ability were demographics (AUC, 0.78, 95% CI, 0.77-0.78), cognitive tests (AUC, 0.72, 95% CI, 0.72-0.73), and IADLs (AUC, 0.71, 95% CI, 0.70-0.71). The most predictive risk factors and the existing models performed inconsistently across socioeconomic groups and had significantly higher AUCs when making predictions for female and uneducated elderly compared to male and educated elderly.

Our study showed that three existing prediction models had significantly lower AUCs when predicting future cognitive impairment among male, urban, and educated elderly Chinese compared to female, rural dwelling, and uneducated elderly. Despite the only shared risk factors in all three models being age and baseline MMSE score, significant differences in predictive ability were consistent across every model. One explanation is risk factor differences between those who developed cognitive impairments and those who did not were larger among the groups where more accurate predictions were made. For example, the difference in average age between women who did and did not become cognitively impaired was 11.8 years, whereas for men it was 7.0 years. Similarly, among the uneducated, the difference in baseline MMSE score between the cognitively normal and cognitively impaired at follow-up was 2.67 compared to 1.45 among the educated. In addition, the prevalence of cognitive impairments in our sample was higher among female, rural dwelling, and uneducated elderly, meaning the models had more events to learn from. Indeed, previous studies using nationally representative data have also reported higher prevalence estimates among these groups[23]. Our results indicate that targeted prediction models for specific socioeconomic groups are needed in China to make equitably accurate risk assessments across sex, residential status, and education level. Several studies have called for such models[24,25], but as of this writing none have been developed in China.

Out of nine risk factor groups, we found that demographics, cognitive tests, and instrumental activities of daily living (IADLs) best predict future cognitive impairment in the general Chinese population and across sex, residential status, and education level. Demographics are often included in prediction models for cognitive impairments[24,26-29], and we suggest they continue to be leveraged because of their predictive power and ease to collect. Associations between chronic diseases, activities of daily living, psychological factors, and diet with cognitive impairments among elderly Chinese have been established[30-37], but such factors showed moderate predictive ability in our study. To our knowledge dietary factors have not been incorporated into existing prediction models in China, but they had higher AUCs than commonly used risk factors such as psychological variables, ADLs, and chronic diseases. In fact, chronic diseases did not make significantly better than random predictions among men (AUC, 0.50, 95% CI, 0.49-0.51), urban (AUC, 0.50, 95% CI, 0.50-0.51), and educated elderly (AUC, 0.51, 95% CI 0.49-0.52). Hence, in addition to providing a ranking of the most predictive risk factor groups, our study is the first to show that dietary factors warrant consideration when predicting future cognitive impairment among elderly Chinese.

Many risk factor groups had significantly different AUCs across population subsets. Like the existing models we re-created, our study revealed that the most predictive risk factors (demographics, cognitive tests, and IADLs) had significantly higher AUCs when making predictions among female and uneducated elderly compared to male and educated elderly. As was the case with the re-created models, this likely resulted from the distributions of risk factors being more separable between those who developed cognitive impairments and those who did not in the groups where more accurate predictions were made. Given the lack of available evidence, it is unclear whether the discrepancies in predictive ability found in our study generalize outside of China, and future work may seek to perform similar analyses elsewhere.

Our study has several limitations. Due to insufficient reporting we cannot guarantee the models we re-created were identical to the original versions. However, the AUCs of each model in the general population were consistent with the reported AUCs in the original papers. Second, the CLHLS is not nationally representative, though it does include elderly from 23 of China's provinces. Lastly, the exercise and sleep group did not include objective measurements of physical activity and sleep. Self-reported exercise and sleep are often inaccurate, and we suggest that the results be interpreted with caution for the exercise and sleep group.

## Conclusions

Out of nine risk factor groups our study found that demographics, cognitive tests, and IADLs best predicted future cognitive impairment among elderly Chinese and had significantly better predictive ability among female and uneducated elderly compared to male and educated elderly. Similarly, every existing model we re-created made significantly better predictions among women, the uneducated, and rural dwelling elderly. Our study suggests more targeted predictions models for cognitive impairment are needed to make equally accurate risk assessments across different socioeconomic groups in China and provides foundational evidence that can support variable selection for such models.

## Acknowledgement

This project was funded by the City University of Hong Kong, Hong Kong SAR, China internal research grant #9610473. C.Sakal and X.Li had full access to all the data in the study and take responsibility for the integrity of the data and the accuracy of the data analysis.

# References

1.	Fang EF, Scheibye-Knudsen M, Jahn HJ, et al. A research agenda for aging in China in the 21st century. *Ageing Res Rev*. Nov 2015;24(Pt B):197-205. doi:10.1016/j.arr.2015.08.003
2.	Fang EF, Xie C, Schenkel JA, et al. A research agenda for ageing in China in the 21st century (2nd edition): Focusing on basic and translational research, long-term care, policy and social networks. *Ageing Res Rev*. Dec 2020;64:101174. doi:10.1016/j.arr.2020.101174
3.	Jia L, Du Y, Chu L, et al. Prevalence, risk factors, and management of dementia and mild cognitive impairment in adults aged 60 years or older in China: a cross-sectional study. *The Lancet Public Health*. 2020;5(12):e661-e671. doi:10.1016/s2468-2667(20)30185-7
4.	Gao F, Lv X, Dai L, et al. A combination model of AD biomarkers revealed by machine learning precisely predicts Alzheimer's dementia: China Aging and Neurodegenerative Initiative (CANDI) study. *Alzheimer's & Dementia*. 2022;doi:10.1002/alz.12700
5.	Hu M, Shu X, Yu G, Wu X, Valimaki M, Feng H. A Risk Prediction Model Based on Machine Learning for Cognitive Impairment Among Chinese Community-Dwelling Elderly People With Normal Cognition: Development and Validation Study. *J Med Internet Res*. Feb 24 2021;23(2):e20298. doi:10.2196/20298
6.	Pu L, Pan D, Wang H, et al. A predictive model for the risk of cognitive impairment in community middle-aged and older adults. *Asian J Psychiatr*. Jan 2023;79:103380. doi:10.1016/j.ajp.2022.103380
7.	Wang S, Wang W, Li X, et al. Using machine learning algorithms for predicting cognitive impairment and identifying modifiable factors among Chinese elderly people. *Front Aging Neurosci*. 2022;14:977034. doi:10.3389/fnagi.2022.977034
8.	Zhou J, Lv Y, Mao C, et al. Development and Validation of a Nomogram for Predicting the 6-Year Risk of Cognitive Impairment Among Chinese Older Adults. *J Am Med Dir Assoc*. Jun 2020;21(6):864-871 e6. doi:10.1016/j.jamda.2020.03.032
9.	Livingston G, Huntley J, Sommerlad A, et al. Dementia prevention, intervention, and care: 2020 report of the Lancet Commission. *The Lancet*. 2020;396(10248):413-446. doi:10.1016/s0140-6736(20)30367-6
10.	Pepe MS, Janes H, Longton G, Leisenring W, Newcomb P. Limitations of the odds ratio in gauging the performance of a diagnostic, prognostic, or screening marker. *Am J Epidemiol*. May 1 2004;159(9):882-90. doi:10.1093/aje/kwh101
11.	Liu D, Li L, An L, et al. Urban–rural disparities in mild cognitive impairment and its functional subtypes among community-dwelling older residents in central China. *General Psychiatry*. 2021;34(5):e100564. doi:10.1136/gpsych-2021-100564
12.	Ren R, Qi J, Lin S, et al. The China Alzheimer Report 2022. *General Psychiatry*. 2022;35(1):e100751. doi:10.1136/gpsych-2022-100751
13.	Ding D, Zhao Q, Wu W, et al. Prevalence and incidence of dementia in an older Chinese population over two decades: The role of education. *Alzheimer's & Dementia*. 2020;16(12):1650-1662. doi:10.1002/alz.12159
14.	Zeng Y. Towards Deeper Research and Better Policy for Healthy Aging --Using the Unique Data of Chinese Longitudinal Healthy Longevity Survey. *China Economic J*. 2012;5(2-3):131-149. doi:10.1080/17538963.2013.764677
15.	Yi Z. Introduction to the Chinese Longitudinal Healthy Longevity Survey (CLHLS). Springer Netherlands; 2008:23-38.
16.	Riley RD, Snell KI, Ensor J, et al. Minimum sample size for developing a multivariable prediction model: PART II - binary and time-to-event outcomes. *Statistics in Medicine*. 2019;38(7):1276-1296. doi:10.1002/sim.7992

17. Collins GS, Reitsma JB, Altman DG, Moons K. Transparent reporting of a multivariable prediction model for individual prognosis or diagnosis (TRIPOD): the TRIPOD Statement. *BMC Medicine*. 2015;13(1):1. doi:10.1186/s12916-014-0241-z
18. Moons KG, Altman DG, Reitsma JB, et al. Transparent Reporting of a multivariable prediction model for Individual Prognosis or Diagnosis (TRIPOD): explanation and elaboration. *Ann Intern Med*. Jan 6 2015;162(1):W1-73. doi:10.7326/M14-0698
19. Katzman R, Zhang MY, Ouang Ya Q, et al. A Chinese version of the Mini-Mental State Examination; impact of illiteracy in a Shanghai dementia survey. *J Clin Epidemiol*. 1988;41(10):971-8. doi:10.1016/0895-4356(88)90034-0
20. Li H, Jia J, Yang Z. Mini-Mental State Examination in Elderly Chinese: A Population-Based Normative Study. *J Alzheimers Dis*. May 7 2016;53(2):487-96. doi:10.3233/JAD-160119
21. Smith GCS, Seaman SR, Wood AM, Royston P, White IR. Correcting for Optimistic Prediction in Small Data Sets. *American Journal of Epidemiology*. 2014;180(3):318-324. doi:10.1093/aje/kwu140
22. Beretta L, Santaniello A. Nearest neighbor imputation algorithms: a critical evaluation. *BMC Medical Informatics and Decision Making*. 2016;16(S3)doi:10.1186/s12911-016-0318-z
23. Jia L, Du Y, Chu L, et al. Prevalence, risk factors, and management of dementia and mild cognitive impairment in adults aged 60 years or older in China: a cross-sectional study. *Lancet Public Health*. Dec 2020;5(12):e661-e671. doi:10.1016/S2468-2667(20)30185-7
24. Hou XH, Feng L, Zhang C, Cao XP, Tan L, Yu JT. Models for predicting risk of dementia: a systematic review. *J Neurol Neurosurg Psychiatry*. Apr 2019;90(4):373-379. doi:10.1136/jnnp-2018-318212
25. Chen Y, Qian X, Zhang Y, et al. Prediction Models for Conversion From Mild Cognitive Impairment to Alzheimer's Disease: A Systematic Review and Meta-Analysis. *Front Aging Neurosci*. 2022;14:840386. doi:10.3389/fnagi.2022.840386
26. Stephan BCM, Pakpahan E, Siervo M, et al. Prediction of dementia risk in low-income and middle-income countries (the 10/66 Study): an independent external validation of existing models. *Lancet Glob Health*. Apr 2020;8(4):e524-e535. doi:10.1016/S2214-109X(20)30062-0
27. Calvin CM, Wilkinson T, Starr JM, et al. Predicting incident dementia 3-8 years after brief cognitive tests in the UK Biobank prospective study of 500,000 people. *Alzheimers Dement*. Dec 2019;15(12):1546-1557. doi:10.1016/j.jalz.2019.07.014
28. Stephan BCM, Pakpahan E, Siervo M, et al. Prediction of dementia risk in low-income and middle-income countries (the 10/66 Study): an independent external validation of existing models. *The Lancet Global Health*. 2020;8(4):e524-e535. doi:10.1016/s2214-109x(20)30062-0
29. You J, Zhang Y-R, Wang H-F, et al. Development of a novel dementia risk prediction model in the general population: A large, longitudinal, population-based machine-learning study. *eClinicalMedicine*. 2022;53:101665. doi:10.1016/j.eclinm.2022.101665
30. Xiao S, Shi L, Zhang J, et al. The role of anxiety and depressive symptoms in mediating the relationship between subjective sleep quality and cognitive function among older adults in China. *J Affect Disord*. Mar 15 2023;325:640-646. doi:10.1016/j.jad.2023.01.048
31. Zhou S, Wang Q, Zhang J, et al. Depressive Symptoms and Cognitive Decline Among Chinese Rural Elderly Individuals: A Longitudinal Study With 2-Year Follow-Up. *Front Public Health*. 2022;10:939150. doi:10.3389/fpubh.2022.939150
32. Li XX, Li Z. The impact of anxiety on the progression of mild cognitive impairment to dementia in Chinese and English data bases: a systematic review and meta-analysis. *Int J Geriatr Psychiatry*. Jan 2018;33(1):131-140. doi:10.1002/gps.4694
33. Guo M, Kang K, Wang A, et al. Association of diabetes status with cognitive impairment in two Chinese rural communities. *J Neurol Sci*. Aug 15 2020;415:116894. doi:10.1016/j.jns.2020.116894
34. Lv YB, Zhu PF, Yin ZX, et al. A U-shaped Association Between Blood Pressure and Cognitive Impairment in Chinese Elderly. *J Am Med Dir Assoc*. Feb 1 2017;18(2):193 e7-193 e13. doi:10.1016/j.jamda.2016.11.011

35. Zhao Z, Li B, Wei C, Sha F. Alcohol consumption and cognitive impairment among Chinese older adults: 8-year follow-up of Chinese Longitudinal Healthy Longevity Study. *Alzheimer's & Dementia*. 2021;17(S10)doi:10.1002/alz.057651
36. Jia J, Zhao T, Liu Z, et al. Association between healthy lifestyle and memory decline in older adults: 10 year, population based, prospective cohort study. *BMJ*. Jan 25 2023;380:e072691. doi:10.1136/bmj-2022-072691
37. Lee ATC, Richards M, Chan WC, Chiu HFK, Lee RSY, Lam LCW. Lower risk of incident dementia among Chinese older adults having three servings of vegetables and two servings of fruits a day. *Age Ageing*. Sep 1 2017;46(5):773-779. doi:10.1093/ageing/afx018

# Figures

**Table 1.** Cohort Characteristics

|  | No. (%) | | |
| --- | --- | --- | --- |
| Characteristic |  | Developed cognitive impairments | |
|  | Everyone (n = 4047) | Yes (n = 337) | No (n = 3710) |
| Age, mean (SD), y | 79.8 (9.4) | 89.1 (9.8) | 79.0 (8.9) |
| Sex |  |  |  |
|   Male | 2037 (50.3) | 130 (38.6) | 1907 (51.4) |
|   Female | 2010 (49.7) | 207 (61.4) | 1803 (48.6) |
| Years of schooling, mean (SD), y | 2.8 (3.7) | 1.8 (3.2) | 2.9 (3.7) |
| Household income, mean (SD), CNY | 24483.8 (25778.6) | 22942.1 (23198.5) | 24623.1 (25997.7) |
| Marital Status |  |  |  |
|   Married and living with spouse | 2033 (50.3) | 83 (24.6) | 1950 (52.7) |
|   Married but not living with spouse | 89 (2.2) | 4 (1.2) | 85 (2.3) |
|   Divorced | 8 (0.2) | 0 (0) | 8 (0.2) |
|   Widowed | 1862 (46.1) | 246 (73) | 1616 (43.7) |
|   Never married | 46 (1.1) | 4 (1.2) | 42 (1.1) |
| Residential Status |  |  |  |
|   City | 665 (16.4) | 57 (16.9) | 608 (16.4) |
|   Town | 1241 (30.7) | 89 (26.4) | 1152 (31.1) |
|   Rural | 2141 (52.9) | 191 (56.7) | 1950 (52.6) |
| Baseline MMSE, mean (SD) | 27.5 (2.8) | 25.1 (3.6) | 27.7 (2.6) |
| Follow-up MMSE, mean (SD) | 26.2 (5.2) | 12.8 (6.1) | 27.5 (2.8) |

**Figure 1. Average** AUC by predictor group and target population

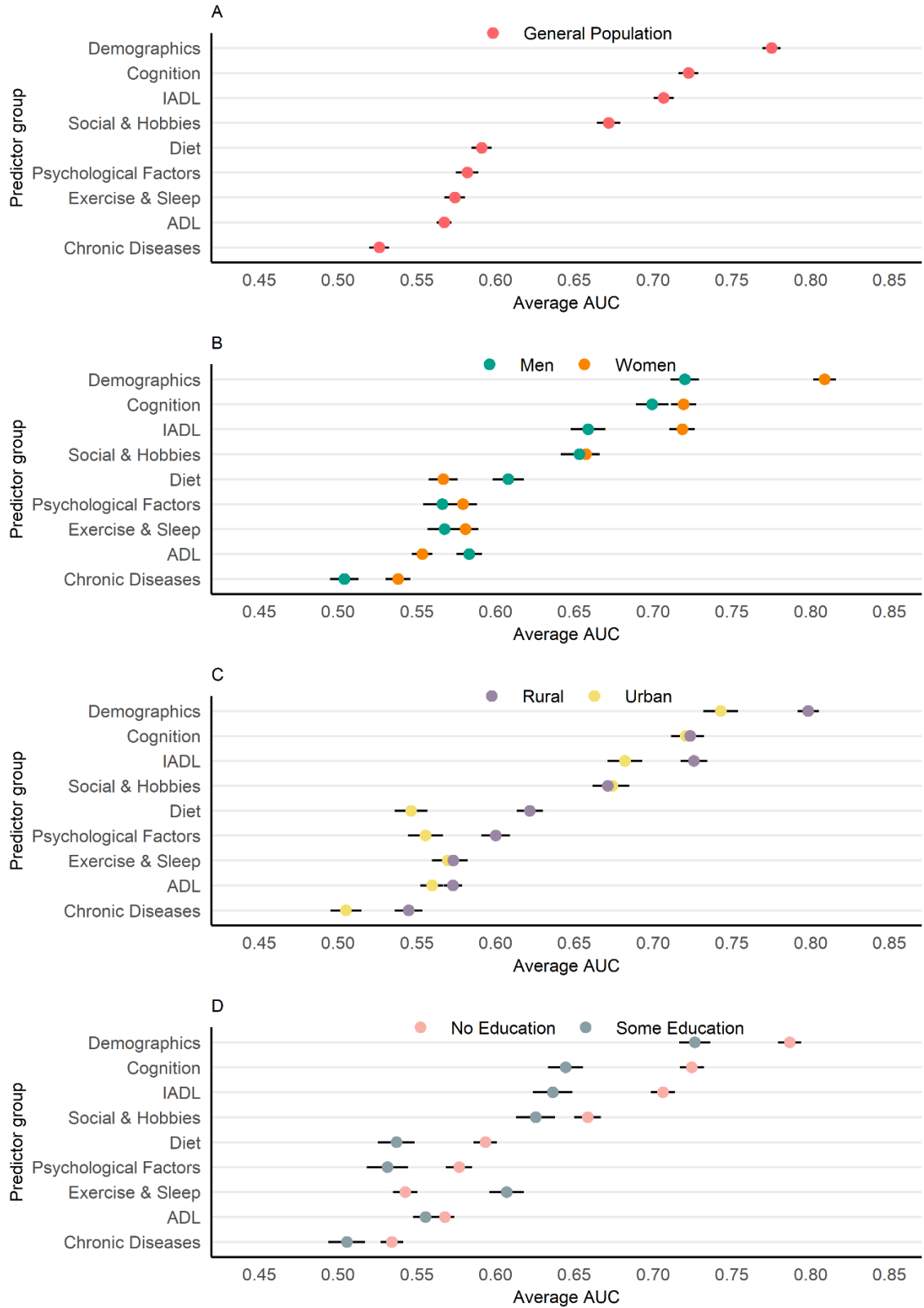

**Table 2.** Predictive ability by target population

| Group/Model | AUC (95% CI) Target Population | | | | | | |
|---|---|---|---|---|---|---|---|
| | General population | Men | Women | Rural | Urban | Uneducated | Educated |
| Demographics | 0.78 (0.77, 0.78) | 0.72 (0.71, 0.73) | 0.81 (0.80, 0.82) | 0.80 (0.79, 0.81) | 0.74 (0.73, 0.75) | 0.79 (0.78, 0.79) | 0.73 (0.72, 0.74) |
| Cognition | 0.72 (0.72, 0.73) | 0.70 (0.69, 0.71) | 0.72 (0.71, 0.73) | 0.72 (0.71, 0.73) | 0.72 (0.71, 0.73) | 0.72 (0.72, 0.73) | 0.64 (0.63, 0.66) |
| IADLs | 0.71 (0.70, 0.71) | 0.66 (0.65, 0.67) | 0.72 (0.71, 0.73) | 0.73 (0.72, 0.73) | 0.68 (0.67, 0.69) | 0.71 (0.70, 0.71) | 0.64 (0.62, 0.65) |
| Social & hobbies | 0.67 (0.66, 0.68) | 0.65 (0.64, 0.66) | 0.66 (0.65, 0.67) | 0.67 (0.66, 0.68) | 0.67 (0.66, 0.68) | 0.66 (0.65, 0.67) | 0.63 (0.61, 0.64) |
| Diet | 0.59 (0.58, 0.60) | 0.61 (0.60, 0.62) | 0.57 (0.56, 0.58) | 0.62 (0.61, 0.63) | 0.55 (0.54, 0.56) | 0.59 (0.59, 0.60) | 0.54 (0.53, 0.55) |
| Psychological factors | 0.58 (0.57, 0.59) | 0.57 (0.55, 0.58) | 0.58 (0.57, 0.59) | 0.60 (0.59, 0.61) | 0.56 (0.54, 0.57) | 0.58 (0.57, 0.59) | 0.53 (0.52, 0.54) |
| Exercise and sleep | 0.57 (0.57, 0.58) | 0.57 (0.56, 0.58) | 0.58 (0.57, 0.59) | 0.57 (0.56, 0.58) | 0.57 (0.56, 0.58) | 0.54 (0.53, 0.55) | 0.61 (0.60, 0.62) |
| ADLs | 0.57 (0.56, 0.57) | 0.58 (0.58, 0.59) | 0.55 (0.55, 0.56) | 0.57 (0.57, 0.58) | 0.56 (0.55, 0.57) | 0.57 (0.56, 0.57) | 0.56 (0.55, 0.56) |
| Chronic Diseases | 0.53 (0.52, 0.53) | 0.50 (0.49, 0.51) | 0.54 (0.53, 0.55) | 0.54 (0.54, 0.55) | 0.50 (0.50, 0.51) | 0.53 (0.53, 0.54) | 0.51 (0.49, 0.52) |
| Wang (2022) | 0.80 (0.80, 0.81) | 0.78 (0.77, 0.78) | 0.82 (0.81, 0.82) | 0.82 (0.81, 0.83) | 0.78 (0.77, 0.79) | 0.81 (0.81, 0.82) | 0.75 (0.74, 0.76) |
| Zhou (2021) | 0.80 (0.80, 0.81) | 0.78 (0.77, 0.79) | 0.82 (0.81, 0.82) | 0.82 (0.81, 0.83) | 0.78 (0.77, 0.79) | 0.82 (0.81, 0.83) | 0.75 (0.74, 0.76) |
| Hu (2020) | 0.80 (0.80, 0.81) | 0.77 (0.77, 0.78) | 0.82 (0.81, 0.83) | 0.82 (0.81, 0.83) | 0.78 (0.77, 0.79) | 0.82 (0.81, 0.82) | 0.75 (0.74, 0.76) |

**Figure 2.** Sensitivity and specificity curves for predictions made in the general population

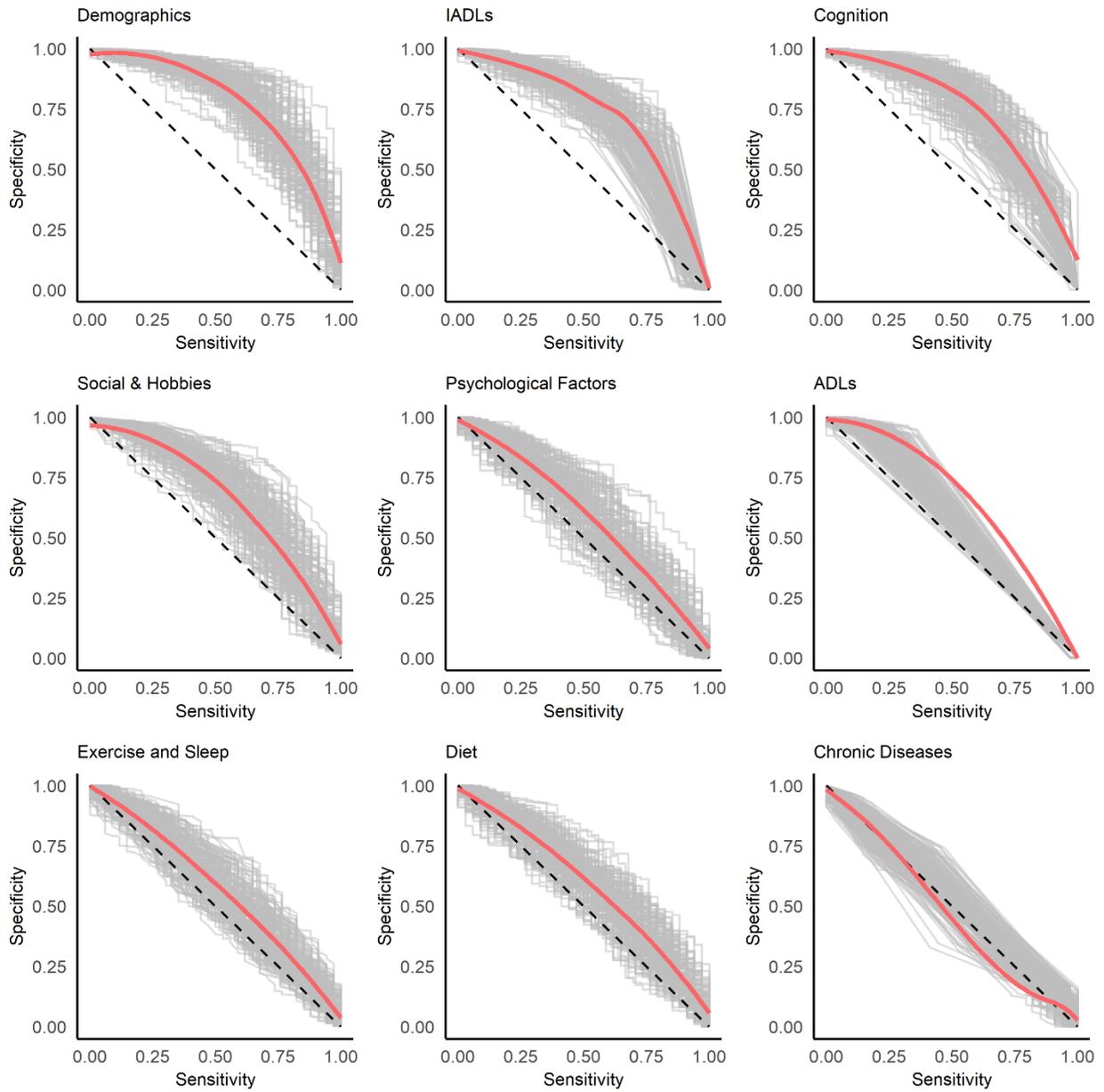

**Figure 3.** Predictive ability of existing models

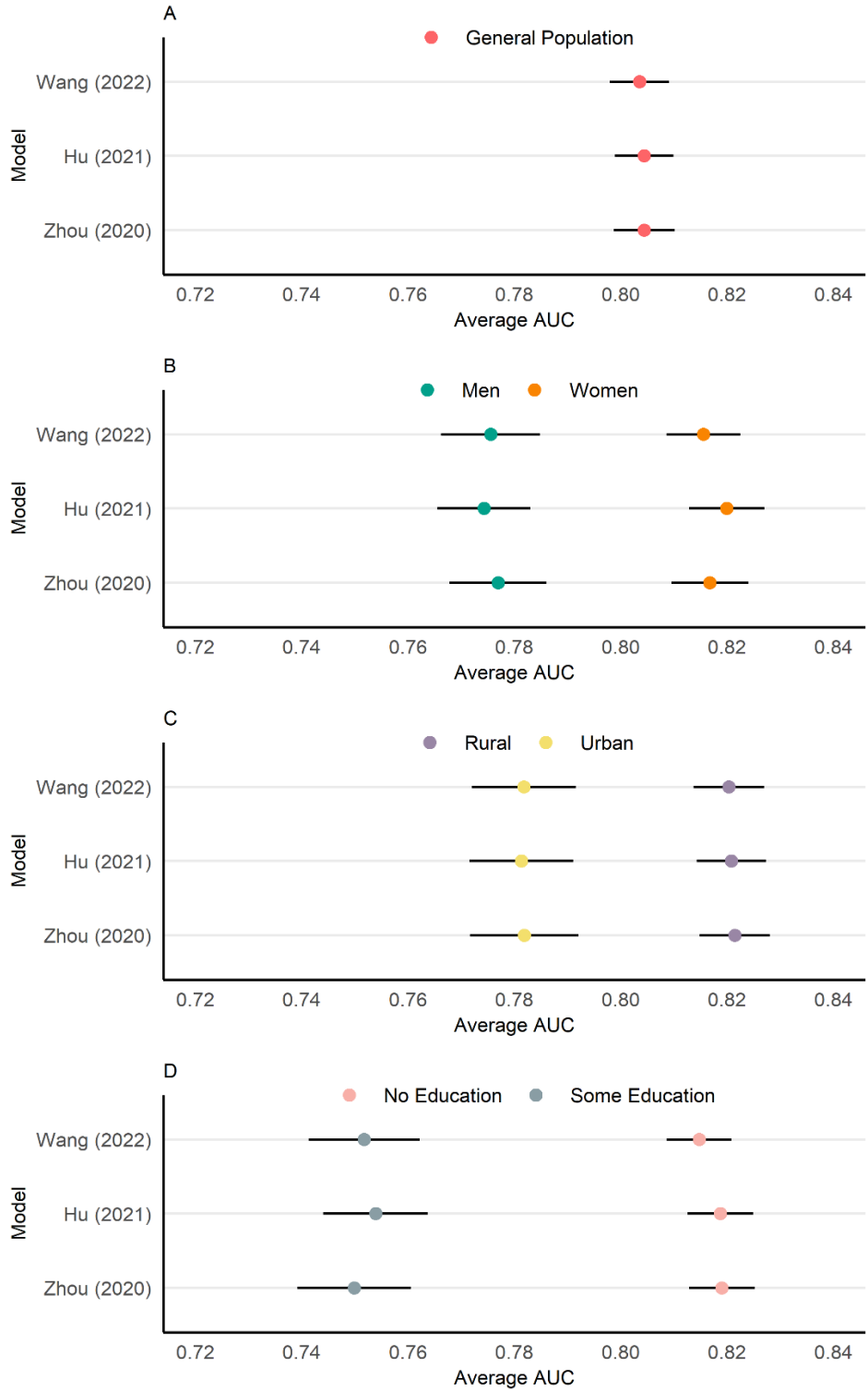